\documentclass{Interspeech2024}
\usepackage{epsfig,amssymb,amsmath,bm}
\usepackage{comment}

\interspeechcameraready

\title{Challenging margin-based speaker embedding extractors by using the variational information bottleneck}

\name[affiliation={1,2}]{Themos}{Stafylakis}
\name[affiliation={3}]{Anna}{Silnova}
\name[affiliation={3}]{Johan}{Rohdin}
\name[affiliation={3}]{Oldrich}{Plchot}
\name[affiliation={3}]{Lukas}{Burget}

\address{
  $^1$Athens University of Economics and Business, Greece \\
  $^2$Omilia - Conversational Intelligence, Athens, Greece \\
  $^3$Brno University of Technology, Speech@FIT and IT4I Center of Excellence, Brno, Czechia 
\email{tstafylakis@aueb.gr, isilnova@fit.vutbr.cz}}


\begin{document}

\maketitle

\begin{abstract}
    Speaker embedding extractors are typically trained using a classification loss over the training speakers. During the last few years, the standard softmax/cross-entropy loss has been replaced by the margin-based losses, yielding significant improvements in speaker recognition accuracy. Motivated by the fact that the margin merely reduces the logit of the target speaker during training, we consider a probabilistic framework that has a similar effect. The variational information bottleneck provides a principled mechanism for making deterministic nodes stochastic, resulting in an implicit reduction of the posterior of the target speaker. We experiment with a wide range of speaker recognition benchmarks and scoring methods and report competitive results to those obtained with the state-of-the-art Additive Angular Margin loss.
\end{abstract}

\section{Introduction}

During the last several years, speaker embeddings extracted with speaker-discriminatively trained deep neural networks have attained impressive performance on several datasets. During this period, the field has witnessed numerous architectures, such as LSTM-based models (\cite{heigold2016end}), 1D (TDNN, ECAPA-TDNN, xi-vectors, TitaNet \cite{snyder2018x,desplanques2020ecapa,lee2021xi,koluguri2022titanet}) and 2D CNNs (ResNet, ResNeXt, Res2Net \cite{zeinali2019but,rybicka2021spine2net,zhou2021resnext}), and more recently self-supervised pretrained Transformer models (Wav2Vec 2.0, HuBERT, WavLM \cite{novoselov2022robust,peng2023parameter}), fined-tuned for the particular downstream task. 

One of the most effective and architecture-agnostic methods to improve speaker embedding discriminability has been the margin-based loss, especially the Additive-Angular Margin (AAM \cite{deng2019arcface,xiang2019margin}) variant. Originating from face recognition, AAM combines directional statistics with a margin penalty on the angle between the embedding and the prototype of the target class. It comes as a drop-in replacement of regular softmax/cross-entropy loss and achieves substantial improvements in many speaker recognition benchmarks. 

On the other hand, the effectiveness of margin in other, more challenging datasets is questionable. Margin-based losses do not retain much of their strength when a backend model (e.g., probabilistic linear discriminant analysis, PLDA) needs to be introduced in the pipeline \cite{BUTOdyssey22,villalba2020state}. This is crucial since most of the successful systems in NIST-SRE do make use of such a backend. Furthermore, in industry-level voice biometrics, where a single extractor typically serves several different deployments (often in different languages), a backend model trained on in-domain data is usually compulsory for attaining state-of-the-art performance. Finally, margin-based losses are not based on a probabilistic framework, which is a desired property for tasks and methods like calibration, self-supervised training, and multitask learning. 

In this paper, we propose an alternative to margin-based loss, based on the variational information bottleneck (VIB \cite{alemi2016deep}). The method has been introduced in deep learning for some years, and although it has been applied to many fields (e.g., speech antispoofing, computer vision, NLP, explainability, a.o. \cite{eom2022anti,du2020learning,cui2024enhancing,gu2020sentiment,bang2021explaining}), it has not been sufficiently explored for training speaker embedding extractors. In \cite{wang2021variational}, an approach similar to ours is presented, however, the experiments are conducted only with VoxCeleb1, which is considered low-resource, with limited test-set, and having low speaker variability compared to, e.g., NIST-SRE or CNCeleb. Furthermore, to the best of our knowledge, we are the first to underscore the similarities between VIB and margin-based losses, conduct experiments on challenging benchmarks such as NIST-SRE, and combine it with backend models (e.g., PLDA).

In the rest of the paper, we examine whether making the embedding stochastic and sampling from its conditional distribution during training may have a similar impact on the speaker-discriminability of the embeddings. We experiment with a wide collection of speaker recognition benchmarks, namely VoxCeleb~\cite{Nagrani19,chung2018voxceleb2}, CNCeleb~\cite{li2022cn}, and the latest NIST-SREs~\cite{nist_sre16,nist_sre18,nist_sre21}. Moreover, we examine different backend methods, ranging from pure cosine-scoring to PLDA and Toroidal Probabilistic Spherical Discriminant Analysis (T-PSDA)~\cite{silnova2023toroidal}. Finally, we provide arguments in favor of further exploration of VIB and propose future research directions. 

\section{Training the network with VIB}
In this section, the method for training the embedding extractor with VIB is demonstrated. We begin by providing some intuition of the Information Bottleneck principle and the derivation of VIB. We then discuss similarities of VIB with other methods, while providing implementation details of our method. 

\subsection{The Information Bottleneck principle}
The Information Bottleneck~\cite{tishby2000information} builds on the desired properties of a model utilizing some internal representation: such representation has to be effective at performing the task the model is used for; at the same time, the information about the input data contained in such representation should be as compressed as possible. In IB, these general considerations are formalized by introducing the following maximization problem:

\begin{equation}
\label{eq:ib}
R_{IB} = I(Z,Y;\bm{\theta}) - \beta I(Z,X;\bm{\theta}),
\end{equation}
where the second term corresponds to the mutual information between the input $X$ and its internal representation $Z$ that should be minimized, and the first one is the mutual information between $Z$ and the output $Y$ to be maximized. The scalar $\beta>0$ controls the amount of compression of $X$ in $Z$, while $\bm{\theta}$ is a vector of model parameters.





\subsection{Derivation of VIB}
\label{VIBDeriv}
In practice, for an arbitrary model, estimating mutual information terms in eq.~\eqref{eq:ib} can be challenging. Deep Variational Information Bottleneck (VIB)~\cite{alemi2016deep} addresses this issue by introducing two variational approximations to the IB by parameterizing it as a neural network. The first is $q(y|z;\bm{\psi})$, which is an approximation to the true conditional distribution $p(y|z)$, where the former is parametrized by the decoder's parameters $\bm{\psi}$, which in our case corresponds to the linear layer that maps $Z$ to the logits (i.e., it is a collection of speaker prototypes, one for each training speaker). The second variational approximation is $r(z)$, with which we approximate the marginal distribution of $Z$, i.e., $r(z) \approx p(z) = \int p(z|x;\bm{\theta})p(x) dx$. Although the parameters of $r(z)$ can be learnable, it is common to keep them fixed and equal to those of a standardized multivariate normal (MVN), i.e., $r(z) = {\cal N}(z|0,I)$ -- the same is used in the experiments of this paper. 

Using the above variational approximations, we obtain 
\begin{equation}
I(Z,Y;\bm{\theta}) \geq \int p(x)p(y|x)p(z|x;\bm{\theta})\log q(y|z;\bm{\psi})dxdydz 
\end{equation}
which is equal to the negative expected cross-entropy, where the expectation is with respect to the training data $p(x)p(y|x)$ but also with respect to $p(z|x;\bm{\theta})$.

Similarly, $I(Z,X;\bm{\theta})$ is upper-bounded by the following expression
\begin{equation}
I(Z,X;\bm{\theta}) \leq \int p(x)p(z|x;\bm{\theta})\log \frac{p(z|x;\bm{\theta})}{r(z)} dxdz 
\end{equation}
which is equal to the expected Kullback-Leibler (KL) divergence between $p(z|x;\bm{\theta})$ and $r(z)$, with respect to the training data distribution $p(x)$.
Therefore, $R_{IB}$ is lower-bounded by the following expression
\begin{multline}
\label{eq:RIB}
R_{IB} \geq \mathbb{E}_{p(x)p(y|x)p(z|x;\bm{\theta})} [\log q(y|z;\bm{\psi})] \\
- \beta \mathbb{E}_{p(x)p(z|x;\bm{\theta})}\left[\log \frac{p(z|x;\bm{\theta})}{r(z)}\right] \\
= 
- \mathbb{E}_{p(x)p(z|x;\bm{\theta})} [\text{CE}(p(y|x),q(y|z;\bm{\psi})] \\
- \beta \mathbb{E}_{p(x)} [\text{KL}(p(z|x;\bm{\theta}),r(z))] \\
\approx \frac{1}{n} \sum_{i=1}^n \bigg[\frac{1}{m}\sum_{j=1}^m \left(\log q(y_{i}|z^{(j)};\bm{\psi})\right) \\
- \beta \text{KL}(p(z|x_i;\bm{\theta}),r(z))\bigg]
\end{multline} 

\subsection{Sampling from the conditional distribution via the reparametrization trick}
Similarly to variational autoencoders (VAEs \cite{kingma2014auto}), VIB models $p(z|x;\bm{\theta})$ by a MVN with diagonal covariance matrix ${\cal N}(z|f_{\mu}(x),\text{diag}(f_{\sigma}^2(x)))$. Therefore, the KL term in eq. (\ref{eq:RIB}) for a given $x$ is simply the KL divergence between two MVN distributions. For calculating the CE term in eq. (\ref{eq:RIB}) for each given $x$, samples from $p(z|x;\bm{\theta})$ are required. The reparametrization trick of VAEs is utilized, where samples $e^{(j)}$ are generated from a standardized MVN and transformed to samples $z^{(j)}$ from the target distribution $p(z|x;\theta)$ as follows
\begin{equation}
z^{(j)} = f_{\sigma}(x) \odot e^{(j)} + f_{\mu}(x), \, e^{(j)}\sim {\cal N}(0,I). 
\end{equation}
where $j=1,\ldots,J$. We implement $f_{\mu}(x)$ and $f_{\sigma}(x)$ with two linear layers having the statistics pooling layer as common input. Note that statistics pooling is defined as mean and std pooling for each frequency-channel pair of the feature maps of the last convolutional layer. These mean and std pooled features should not be confused with $f_{\mu}(x)$ and $f_{\sigma}(x)$, as the latter correspond to the mean and std parametrizing the distribution of the stochastic embedding $Z$. The non-negativity of the diagonal elements of $f_{\sigma}(x)$ is enforced by the softplus function, although the exponential function may also be applied. Finally, the encoder is a ResNet-34 with statistics pooling layer, although other architectures may be considered.

\subsection{VIB and VAE}
As we observe, VIB and VAE have several similarities, since (a) they both make use of a stochastic internal representation $Z$, (b) they prevent the collapse of $p(z|x;\bm{\theta})$ to a point mass by penalizing the KL divergence between it and $p(z)$, and (c) they make use of the reparametrization trick to sample from $p(z|x)$. From this perspective, one may consider VIB as a supervised learning analog of VAEs. On the other hand, there are certain differences, the main of which are the following. 
\begin{itemize}
\item The classifier-decoder of the VIB, as opposed to the (deep) reconstruction-based decoder of VAE. Note that this allows generating many samples $z^{(j)}$ (i.e., $J\gg1$) per training example to reduce the variance of the estimator with a minor increase in computation since the classifier is merely a linear layer (we set $J=10$). 
\item The information-theoretic formulation of VIB, as opposed to the variational-Bayes (VB) formulation of VAEs, which among others provides a better justification of the tunable trade-off parameter $\beta>0$ (The VB formulation implies $\beta=1$). 
\end{itemize}
 
\subsection{Similarities with margin-based losses}
\label{AAM}
Apart from VAEs, VIB has certain similarities with margin-based losses. The AAM loss is a variant of Softmax/CE, where the logit of the $i$th training example and $j$th training speaker is modified as $l_{i,j} = s\cos(\theta_{z_i,y_j} + m\delta_{j,y_i})$, where $s$ is the scale (typically $s=30$), $m$ is the margin, $\delta_{\cdot,\cdot}$ is the Dirac function, and $\theta_{z_i,y_j}$ is the (positive) angle between the embedding $z_i$ and the $j$th speaker prototype. It therefore makes the classification task artificially harder, by intervening to the logits and reducing the one corresponding to the target class. This intervention creates a safety zone around the class boundaries, which increases the discriminability of the speaker embeddings, especially for speakers not appearing in the training set. 

In the VIB framework, $Z$ is assumed stochastic, its conditional distribution $p(z|x;\bm{\theta})$ is estimated and samples are generated from it, which should be classified using regular Softmax/CE. Having to classify samples from $p(z|x;\bm{\theta})$ instead of the expected value $\mathbb{E}_{p(z|x;\bm{\theta})}[Z]$ makes the task harder, as it increases the within-class variability of the embeddings. This results from the fact that $\mathbb{E}_{p(z|x;\bm{\theta})} \log q(y|z;\bm{\psi})\leq \log q(y|\mathbb{E}_{p(z|y;\bm{\psi})} [Z];\bm{\psi})$ where $y$ is the ground-truth speaker. As sampling is applied only during training, it creates a safety zone around the class boundaries, similarly to the AAM loss. 

Note that other regularization techniques are employed for making the task harder during training, such as data augmentation (by adding noise and reverberation or applying SpecAugment), short training utterances, or dropout. Nonetheless, the similarity between VIB and margin stems from the fact that they both act on the classifier of the architecture, while the other methods are either input-level or act on the intermediate layers. Therefore, as with margin-based losses, the other regularization techniques are orthogonal to VIB. 

\subsection{Angular margin, unit-length normalization, and VIB}
Margin-based losses are typically combined with angular distances, e.g., by unit-length normalizing the embedding and the class/speaker prototypes. For example, penalizing the target logit with a fixed additive margin without applying such unit-length normalizations may not be effective, as the network would be free to overcome the penalty by increasing the average magnitude of the embeddings, and therefore increasing the average scale of the logits, rendering the margin ineffective. As shown in Sect.~\ref{AAM}, the AAM unit-length normalizes both the embeddings and speaker prototypes and bounds the target logit to $[-s,s\cos(m)]$.

On the other hand, VIB does not necessitate unit-length normalization of neither the embedding nor the prototypes. As discussed in Sect \ref{VIBDeriv}, the KL term enforces the aggregated conditional distribution $p(z) = \int p(z|x;\bm{\theta})p(x) dx$ of the training set to be as close as possible to $r(z)$ preventing the magnitude (length) of $Z$ from increasing beyond the (soft) boundaries of the support of $r(z)$. This inherent mechanism of VIB provides us with flexibility in choosing whether or not to length-normalize embeddings and/or prototypes, e.g., depending on the benchmark and the scoring method to be used.

\section{Experiments}
We perform three sets of experiments, NIST-SRE, VoxCeleb, and CNCeleb, to analyze the performance of VIB regularization. All of them were conducted using WeSpeaker (\cite{wang2023wespeaker}) toolkit\footnote{https://github.com/wenet-e2e/wespeaker} and closely followed experimental setups of the corresponding recipes including training set, augmentations, and training hyperparameters like optimizer, learning rate, etc. The only exception is that we used longer training examples of 300 frames (instead of the default 200) in all experiments. Also, we use our custom implementation of scoring backends for NIST-SRE experiments. When implementing the training with VIB we reuse the same scheduler as used for the margin in WeSpeaker: for the first 20 epochs, the margin (or $\beta$ in VIB) is set to zero and then exponentially increased to its final value in the course of the next 20 epochs after which it is kept constant for the rest of the training.  
\subsection{NIST-SRE}
In these experiments, we train all embedding extractors on NIST CTS superset~\cite{sre_cts_superset} and test their performance on evaluation sets of three editions of NIST SRE: SRE2016~\cite{nist_sre16}, telephone condition from SRE2018~\cite{nist_sre18}, and audio-only part of SRE 2021~\cite{nist_sre21}.
In all cases, the embeddings were centered and length-normalized. Then, we applied linear discriminant analysis (LDA) reducing the dimensionality of the embeddings from 256 to 100. Finally, we trained two scoring backends on preprocessed embeddings: PLDA and T-PSDA.  When PLDA is used for scoring, both speaker and channel subspaces have a dimensionality of 100 (i.e., we use the two-covariance version of the PLDA model); when T-PSDA is used, we do not attempt to tune its hyperparameters but rather adopt the same values as found optimal in~\cite{silnova2023toroidal}: we use one 60-dimensional speaker variable and two 5-dimensional channel variables. The parameters for centering, LDA, and the backends were estimated on the embedding extractor training set. The performance is reported in terms of equal error rate~(EER) and minimum cost ($\mathrm{min\_C}$) as defined by the respective evaluation plan and computed by the scoring tools provided by the evaluation organizers. 

Table~\ref{tab:results_sre} shows the comparison of the baseline embedding extractors trained with regular Softmax/CE or AAM objectives versus the VIB version of the same objective. In the first column of the table, the number in the brackets corresponds to the value of the margin $m$ for AAM and to $\beta$ for VIB. The table is separated into two parts for two different scoring backends allowing not only comparison between the backends themselves but also showing the effectiveness of the VIB approach across different scoring methods. By analyzing the results, we see that VIB consistently outperforms the corresponding loss not utilizing the margin (VIB vs. CE and VIB\_LN vs. AAM(0.0)) while in most of the cases being competitive to the margin-based AAM(0.2).

\begin{table*}[!htb]
  \caption{Results on NIST SRE evaluation sets.} 
  \label{tab:results_sre}
  \centering
 \begin{tabular}{ c l c c c c  c c c c  }
    \toprule
    & & \multicolumn{2}{c}{\textbf{SRE16 yue}}&\multicolumn{2}{c}{\textbf{SRE16 tgl}}&\multicolumn{2}{c}{\textbf{SRE18 CMN2}}&\multicolumn{2}{c}{\textbf{SRE21 audio}}\\
    \textbf{loss} & \textbf{backend} & 
     $\mathrm{\mathbf{min\_C}}$  & \textbf{EER(\%)}&   $\mathrm{\mathbf{min\_C}}$  & \textbf{EER(\%)}&  $\mathrm{\mathbf{min\_C}}$  & \textbf{EER(\%)} &  $\mathrm{\mathbf{min\_C}}$  & \textbf{EER(\%)} \\
    \midrule	
    CE&PLDA&.470&4.50&.998&20.17&.569&7.81&.564&10.54\\
    VIB(0.002)&PLDA&.326&3.36&.951&14.79&.517&6.56&.555&10.50\\
    AAM(0.0)&PLDA&.336&3.39&.988&15.19&.502&6.55&.552&10.83\\
    AAM(0.2)&PLDA&.336&3.34&\textbf{.849}&\textbf{13.14}&\textbf{.487}&\textbf{6.19}&.562&10.29\\
    VIB\_LN(0.002)&PLDA&\textbf{ .315} &\textbf{3.16} &.984 & 14.91&.491&\textbf{6.19}&\textbf{.538}&\textbf{9.42}\\
    
    \midrule
       CE&T-PSDA&.441&4.80&.872&16.16&.569&7.96&.573&9.98\\
    VIB(0.002)&T-PSDA&.394&4.00&.807&12.65&.541&6.55&\textbf{.565}&10.44\\	
    AAM(0.0)&T-PSDA&\textbf{.355}&3.86&\textbf{.742}&11.37&\textbf{.518}&6.48&.569&10.26\\		
    AAM(0.2)&T-PSDA&.399&3.96&.758&\textbf{10.87}&\textbf{.518}&6.30&.599&10.37\\	
    VIB\_LN(0.002)&T-PSDA&.360&\textbf{3.64}&.779&12.12&.525&\textbf{6.29}&.588&\textbf{9.51}\\	
    \bottomrule
  \end{tabular}

\end{table*}

\subsection{VoxCeleb}
When running the experiments on VoxCeleb dataset, we follow a commonly adopted setup: we use the development part of VoxCeleb2~\cite{chung2018voxceleb2} to train the embedding extractor and the whole VoxCeleb1~\cite{Nagrani19} for testing. As a backend, we use simple cosine scoring with only centering and length-normalization of the embeddings as preprocessing. The centering is done with the mean computed on VoxCeleb2 development set. The same set was used as a cohort for adaptive score normalization, where we selected 300 highest scores from the cohort to estimate normalization parameters. 

Table~\ref{tab:results_vox} displays the results achieved with the baseline extractors trained with AAM with margins 0.2 and 0.0 along with the one that uses VIB. Similar to NIST-SRE case, we observe that VIB consistently outperforms AAM loss with the margin set to zero, although when the margin is used AAM is clearly superior. These observations are valid for both cases: whether we use score-normalization or not.  Apart from this, the table shows the large-margin fine-tuned model along with an analogous setting for VIB: in both cases, the length of the training examples was increased to 6 seconds for the last 10 epochs with the margin increased to 0.5 and the value of $\beta$ kept the same as for the rest of the training (0.004 in this case).

\begin{table*}[!htb]
  \caption{Results on VoxCeleb with cosine scoring, without and with score normalization (wo/w).} 
  \label{tab:results_vox}
  \centering
 \begin{tabular}{ c l  c c  c c c c  }
    \toprule
    && \multicolumn{2}{c}{\textbf{Vox1-O}}&\multicolumn{2}{c}{\textbf{Vox1-E}}&\multicolumn{2}{c}{\textbf{Vox1-H}}\\
     & \textbf{} & 

     $\mathrm{\mathbf{minDCF_{0.01}}}$  & \textbf{EER(\%)}&  $\mathrm{\mathbf{minDCF_{0.01}}}$  & \textbf{EER(\%)} &  $\mathrm{\mathbf{minDCF_{0.01}}}$  & \textbf{EER(\%)} \\
    \midrule	
    AAM(0.0)&&.150/.131 & 1.28/1.11 & .154/.132 & 1.33/1.18 & .245/.192 & 2.55/2.13 \\
    AAM(0.2)&&.115/.090&0.96/0.83 & .114/.104 & 0.98/0.89 & .182/.160 & 1.86/1.63 \\
    	
    VIB\_LN(0.004)&&.109/.094&0.99/0.88&.130/.115&1.11/1.02&.204/.174&2.08/1.82\\	
    \midrule
    AAM(0.2)+FT(0.5)&&.074/.056&0.88/0.74&.101/.092&0.95/0.89&.173/.151&1.69/1.53\\
    VIB\_LN+FT(0.004)&&.113/.078&0.96/0.80&.121/.108&1.06/0.95&.194/.160&1.96/1.71\\	
    \bottomrule
  \end{tabular}

\end{table*}

\subsection{CNCeleb}

Following WeSpeaker CNCeleb recipe, we use a combination of CNCeleb2 and the development set of CNCeleb1 as the training set and evaluate on the test set of CNCeleb1. For multi-enrollment trials, the embeddings for all sessions are extracted and averaged to obtain a single enrollment embedding.

We performed a similar set of experiments to the ones presented for VoxCeleb dataset: we compared the performance of several embedding extractors trained with AAM or VIB with and without finetuning on the long training examples. In all cases, simple cosine scoring was used, with optionally using score normalization with the network training set used as a cohort. The results are given in Table \ref{CNexp} and show that VIB is competitive to AAM loss in most of the metrics.

\begin{table}[!htb]
\centering
\caption{\label{tab:results_cn} Results on the CNCeleb evaluation set, with cosine scoring, without and with score normalization (wo/w).}
\label{CNexp}
\begin{tabular}{cccc}
\toprule
&& $\mathrm{\mathbf{minDCF_{0.01}}}$ & \textbf{EER(\%)}  \\
\midrule
AAM(0.0)      &&.406/.367 & 7.97/7.25\\
AAM(0.2)      &&.394/.360 & 7.22/6.61\\
VIB\_LN(0.004)&&.406/.372 & 7.24/6.87\\
\midrule
AAM(0.2)+FT(0.5) &&.393/.356 & 7.23/6.56\\
VIB\_LN+FT(0.004)&&.399/.364 & 7.29/6.78\\
\bottomrule
\end{tabular}
\end{table}

\subsection{Discussion}
The experiments we conducted on VoxCeleb and CNCeleb show that our implementation of VIB is comparable to the state-of-the-art AAM loss, recovering most of its performance gains compared to AAM with zero margin (i.e., only length-normalization of embeddings and prototypes). The experiments on NIST-SRE show that our implementation of VIB is competitive to the best setting for each evaluation test we examine. 

We emphasize that there are numerous directions one may consider to boost the performance of the method, such as deeper and less mutually-coupled branches for estimating $f_{\mu}(x)$ and $f_{\sigma}(x)$, different $r(z)$ such as Gaussian Mixture Models or von Mises-Fisher distribution, and several optimization techniques that can be found in the rich literature of VIB and VAE. We should also mention that most of the hyperparameters we used for training are optimized for the AAM loss, leaving room for further improvement simply by hyperparameter optimization.

Furthermore, the solid probabilistic/information-theoretic framework of VIB facilitates research in many directions, such as (a) unsupervised domain adaptation and multi-domain training (by adapting or conditioning the marginal $p(z)$ to each domain), (b) uncertainty propagation (by propagating the uncertainly of the conditional distribution $p(z|x;\bm{\theta})$ in the probabilistic backend \cite{kenny2013plda,wang2023incorporating}), (c) disentangled speaker embedding extractors (e.g., \cite{liu2024disentangling}) and finally, (d) incorporating the embedding extractor to more general architectures (e.g., diarization, target-speaker extractor, voice conversion, joint speaker and speech recognition architecture, a.o.). Joint architectures typically behave better with well-defined distributions and losses, and VIB is a step towards this direction.

\section{Conclusions}
In this paper, we examined the strength of VIB in training speaker embedding extractors. Motivated by the wide adoption of margin-based losses, their strength in boosting the performance of virtually any architecture, but also their weaknesses, we tried to address the question of whether or not the VIB can eventually replace margin-based losses. We experimented with a diverse collection of speaker recognition benchmarks and used different scoring methods. Our experiments show that our implementation of VIB yields improvement over margin-based loss in several NIST-SRE test sets, while it can recover much of the performance gains of the margin-based loss in VoxCeleb and CNCeleb. Finally, several directions to improve its performance are provided, together with other speaker-related applications and settings in which VIB may be beneficial.  

\section{Acknowledgements}
The work was partially supported by Czech Ministry of Interior project No. VB02000060, 2024-2026 "NABOSO" and Horizon 2020 Marie Sklodowska-Curie grant ESPERANTO, No. 101007666. Computing on IT4I supercomputer was supported by the Ministry of Education, Youth and Sports of the Czech Republic through e-INFRA CZ (ID:90254).

\bibliographystyle{IEEEtran}
\bibliography{mybib}

\begin{thebibliography}{10}
\providecommand{\url}[1]{#1}
\csname url@samestyle\endcsname
\providecommand{\newblock}{\relax}
\providecommand{\bibinfo}[2]{#2}
\providecommand{\BIBentrySTDinterwordspacing}{\spaceskip=0pt\relax}
\providecommand{\BIBentryALTinterwordstretchfactor}{4}
\providecommand{\BIBentryALTinterwordspacing}{\spaceskip=\fontdimen2\font plus
\BIBentryALTinterwordstretchfactor\fontdimen3\font minus \fontdimen4\font\relax}
\providecommand{\BIBforeignlanguage}[2]{{%
\expandafter\ifx\csname l@#1\endcsname\relax
\typeout{** WARNING: IEEEtran.bst: No hyphenation pattern has been}%
\typeout{** loaded for the language `#1'. Using the pattern for}%
\typeout{** the default language instead.}%
\else
\language=\csname l@#1\endcsname
\fi
#2}}
\providecommand{\BIBdecl}{\relax}
\BIBdecl

\bibitem{heigold2016end}
G.~Heigold, I.~Moreno, S.~Bengio, and N.~Shazeer, ``End-to-end text-dependent speaker verification,'' in \emph{2016 IEEE International Conference on Acoustics, Speech and Signal Processing (ICASSP)}.\hskip 1em plus 0.5em minus 0.4em\relax IEEE, 2016, pp. 5115--5119.

\bibitem{snyder2018x}
D.~Snyder, D.~Garcia-Romero, G.~Sell, D.~Povey, and S.~Khudanpur, ``X-vectors: Robust dnn embeddings for speaker recognition,'' in \emph{2018 IEEE international conference on acoustics, speech and signal processing (ICASSP)}.\hskip 1em plus 0.5em minus 0.4em\relax IEEE, 2018, pp. 5329--5333.

\bibitem{desplanques2020ecapa}
B.~Desplanques, J.~Thienpondt, and K.~Demuynck, ``{ECAPA-TDNN: Emphasized Channel Attention, Propagation and Aggregation in TDNN Based Speaker Verification},'' in \emph{Interspeech}, 2020.

\bibitem{lee2021xi}
K.~A. Lee, Q.~Wang, and T.~Koshinaka, ``Xi-vector embedding for speaker recognition,'' \emph{IEEE Signal Processing Letters}, vol.~28, pp. 1385--1389, 2021.

\bibitem{koluguri2022titanet}
N.~R. Koluguri, T.~Park, and B.~Ginsburg, ``Titanet: Neural model for speaker representation with 1d depth-wise separable convolutions and global context,'' in \emph{ICASSP 2022-2022 IEEE International Conference on Acoustics, Speech and Signal Processing (ICASSP)}.\hskip 1em plus 0.5em minus 0.4em\relax IEEE, 2022, pp. 8102--8106.

\bibitem{zeinali2019but}
H.~Zeinali, S.~Wang, A.~Silnova, P.~Mat{\v{e}}jka, and O.~Plchot, ``But system description to voxceleb speaker recognition challenge 2019,'' \emph{arXiv preprint arXiv:1910.12592}, 2019.

\bibitem{rybicka2021spine2net}
M.~Rybicka, J.~Villalba, P.~Zelasko, N.~Dehak, and K.~Kowalczyk, ``Spine2net: Spinenet with res2net and time-squeeze-and-excitation blocks for speaker recognition.'' in \emph{Interspeech}, 2021, pp. 496--500.

\bibitem{zhou2021resnext}
T.~Zhou, Y.~Zhao, and J.~Wu, ``Resnext and res2net structures for speaker verification,'' in \emph{2021 IEEE Spoken Language Technology Workshop (SLT)}.\hskip 1em plus 0.5em minus 0.4em\relax IEEE, 2021, pp. 301--307.

\bibitem{novoselov2022robust}
S.~Novoselov, G.~Lavrentyeva, A.~Avdeeva, V.~Volokhov, and A.~Gusev, ``Robust speaker recognition with transformers using wav2vec 2.0,'' \emph{arXiv preprint arXiv:2203.15095}, 2022.

\bibitem{peng2023parameter}
J.~Peng, T.~Stafylakis, R.~Gu, O.~Plchot, L.~Mo{\v{s}}ner, L.~Burget, and J.~{\v{C}}ernock{\`y}, ``Parameter-efficient transfer learning of pre-trained transformer models for speaker verification using adapters,'' in \emph{ICASSP 2023-2023 IEEE International Conference on Acoustics, Speech and Signal Processing (ICASSP)}.\hskip 1em plus 0.5em minus 0.4em\relax IEEE, 2023, pp. 1--5.

\bibitem{deng2019arcface}
J.~Deng, J.~Guo, N.~Xue, and S.~Zafeiriou, ``{ArcFace: Additive Angular Margin Loss for Deep Face Recognition},'' in \emph{{Proceedings of the IEEE/CVF Conference on Computer Vision and Pattern Recognition (CVPR)}}, June 2019.

\bibitem{xiang2019margin}
X.~Xiang, S.~Wang, H.~Huang, Y.~Qian, and K.~Yu, ``Margin matters: Towards more discriminative deep neural network embeddings for speaker recognition,'' in \emph{2019 Asia-Pacific Signal and Information Processing Association Annual Summit and Conference (APSIPA ASC)}.\hskip 1em plus 0.5em minus 0.4em\relax IEEE, 2019, pp. 1652--1656.

\bibitem{BUTOdyssey22}
\BIBentryALTinterwordspacing
A.~Silnova, T.~Stafylakis, L.~Mo\v{s}ner, O.~Plchot, J.~Rohdin, P.~Mat\v{e}jka, L.~Burget, O.~Glembek, and N.~Br\"ummer, ``Analyzing speaker verification embedding extractors and back-ends under language and channel mismatch,'' in \emph{Odyssey 2022: The speaker and Language Recongnition Workshop, Beijing}, 2022. [Online]. Available: \url{https://arxiv.org/abs/2203.10300}
\BIBentrySTDinterwordspacing

\bibitem{villalba2020state}
J.~Villalba, N.~Chen, D.~Snyder, D.~Garcia-Romero, A.~McCree, G.~Sell, J.~Borgstrom, L.~P. Garc{\'\i}a-Perera, F.~Richardson, R.~Dehak \emph{et~al.}, ``State-of-the-art speaker recognition with neural network embeddings in nist sre18 and speakers in the wild evaluations,'' \emph{Computer Speech \& Language}, vol.~60, p. 101026, 2020.

\bibitem{alemi2016deep}
A.~A. Alemi, I.~Fischer, J.~V. Dillon, and K.~Murphy, ``Deep variational information bottleneck,'' in \emph{International Conference on Learning Representations}, 2016.

\bibitem{eom2022anti}
Y.~Eom, Y.~Lee, J.~Um, and H.-R. Kim, ``Anti-spoofing using transfer learning with variational information bottleneck,'' in \emph{23rd Annual Conference of the International Speech Communication Association, INTERSPEECH 2022}.\hskip 1em plus 0.5em minus 0.4em\relax ISCA, 2022, pp. 3568--3572.

\bibitem{du2020learning}
Y.~Du, J.~Xu, H.~Xiong, Q.~Qiu, X.~Zhen, C.~G. Snoek, and L.~Shao, ``Learning to learn with variational information bottleneck for domain generalization,'' in \emph{Computer Vision--ECCV 2020: 16th European Conference, Glasgow, UK, August 23--28, 2020, Proceedings, Part X 16}.\hskip 1em plus 0.5em minus 0.4em\relax Springer, 2020, pp. 200--216.

\bibitem{cui2024enhancing}
S.~Cui, J.~Cao, X.~Cong, J.~Sheng, Q.~Li, T.~Liu, and J.~Shi, ``Enhancing multimodal entity and relation extraction with variational information bottleneck,'' \emph{IEEE/ACM Transactions on Audio, Speech, and Language Processing}, 2024.

\bibitem{gu2020sentiment}
T.~Gu, G.~Xu, and J.~Luo, ``Sentiment analysis via deep multichannel neural networks with variational information bottleneck,'' \emph{IEEE Access}, vol.~8, pp. 121\,014--121\,021, 2020.

\bibitem{bang2021explaining}
S.~Bang, P.~Xie, H.~Lee, W.~Wu, and E.~Xing, ``Explaining a black-box by using a deep variational information bottleneck approach,'' in \emph{Proceedings of the AAAI conference on artificial intelligence}, vol.~35, no.~13, 2021, pp. 11\,396--11\,404.

\bibitem{wang2021variational}
D.~Wang, Y.~Dong, Y.~Li, Y.~Zi, Z.~Zhang, X.~Li, and S.~Xiong, ``Variational information bottleneck based regularization for speaker recognition.'' in \emph{Interspeech}, 2021, pp. 1054--1058.

\bibitem{Nagrani19}
A.~Nagrani, J.~S. Chung, W.~Xie, and A.~Zisserman, ``Voxceleb: Large-scale speaker verification in the wild,'' \emph{Computer Speech and Language}, 2019.

\bibitem{chung2018voxceleb2}
J.~Chung, A.~Nagrani, and A.~Zisserman, ``Voxceleb2: Deep speaker recognition,'' \emph{Interspeech 2018}, 2018.

\bibitem{li2022cn}
L.~Li, R.~Liu, J.~Kang, Y.~Fan, H.~Cui, Y.~Cai, R.~Vipperla, T.~F. Zheng, and D.~Wang, ``Cn-celeb: multi-genre speaker recognition,'' \emph{Speech Communication}, vol. 137, pp. 77--91, 2022.

\bibitem{nist_sre16}
S.~O. Sadjadi, T.~Kheyrkhah, A.~Tong, C.~Greenberg, D.~Reynolds, E.~Singer, L.~Mason, and J.~Hernandez-Cordero, ``The 2016 nist speaker recognition evaluation,'' \emph{Interspeech 2017}, 2017.

\bibitem{nist_sre18}
\BIBentryALTinterwordspacing
O.~Sadjadi, C.~Greenberg, E.~Singer, D.~Reynolds, L.~Mason, and J.~Hernandez-Cordero, ``\BIBforeignlanguage{en}{The 2018 nist speaker recognition evaluation}.''\hskip 1em plus 0.5em minus 0.4em\relax INTERSPEECH, Graz, AT, 2019-09-15 00:09:00 2019. [Online]. Available: \url{https://tsapps.nist.gov/publication/get_pdf.cfm?pub_id=927673}
\BIBentrySTDinterwordspacing

\bibitem{nist_sre21}
S.~O. Sadjadi, C.~Greenberg, E.~Singer, L.~Mason, and D.~Reynolds, ``The 2021 nist speaker recognition evaluation,'' \emph{arXiv preprint arXiv:2204.10242}, 2022.

\bibitem{silnova2023toroidal}
A.~Silnova, N.~Br{\"u}mmer, A.~Swart, and L.~Burget, ``Toroidal probabilistic spherical discriminant analysis,'' in \emph{ICASSP 2023-2023 IEEE International Conference on Acoustics, Speech and Signal Processing (ICASSP)}.\hskip 1em plus 0.5em minus 0.4em\relax IEEE, 2023, pp. 1--5.

\bibitem{tishby2000information}
N.~Tishby, F.~C. Pereira, and W.~Bialek, ``The information bottleneck method,'' in \emph{The 37th annual Allerton Conference on Communication, Control, and Computing}, 1999, pp. 368--377.

\bibitem{kingma2014auto}
D.~P. Kingma and M.~Welling, ``Auto-encoding variational bayes,'' in \emph{In Proceedings of the 2nd International Conference on Learning Representations (ICLR)}, 2014.

\bibitem{wang2023wespeaker}
H.~Wang, C.~Liang, S.~Wang, Z.~Chen, B.~Zhang, X.~Xiang, Y.~Deng, and Y.~Qian, ``Wespeaker: A research and production oriented speaker embedding learning toolkit,'' in \emph{ICASSP 2023-2023 IEEE International Conference on Acoustics, Speech and Signal Processing (ICASSP)}.\hskip 1em plus 0.5em minus 0.4em\relax IEEE, 2023, pp. 1--5.

\bibitem{sre_cts_superset}
\BIBentryALTinterwordspacing
O.~Sadjadi, ``\BIBforeignlanguage{en}{{NIST SRE CTS Superset: A large-scale dataset for telephony speaker recognition}},'' 2021-08-16 04:08:00 2021. [Online]. Available: \url{https://tsapps.nist.gov/publication/get_pdf.cfm?pub_id=933116}
\BIBentrySTDinterwordspacing

\bibitem{kenny2013plda}
P.~Kenny, T.~Stafylakis, P.~Ouellet, M.~J. Alam, and P.~Dumouchel, ``Plda for speaker verification with utterances of arbitrary duration,'' in \emph{2013 IEEE International Conference on Acoustics, Speech and Signal Processing}.\hskip 1em plus 0.5em minus 0.4em\relax IEEE, 2013, pp. 7649--7653.

\bibitem{wang2023incorporating}
Q.~Wang, K.~A. Lee, and T.~Liu, ``Incorporating uncertainty from speaker embedding estimation to speaker verification,'' in \emph{ICASSP 2023-2023 IEEE International Conference on Acoustics, Speech and Signal Processing (ICASSP)}.\hskip 1em plus 0.5em minus 0.4em\relax IEEE, 2023, pp. 1--5.

\bibitem{liu2024disentangling}
T.~Liu, K.~A. Lee, Q.~Wang, and H.~Li, ``Disentangling voice and content with self-supervision for speaker recognition,'' \emph{Advances in Neural Information Processing Systems}, vol.~36, 2024.

\end{thebibliography}

\end{document}